\documentclass[a4paper,11pt]{article}
\usepackage{amsfonts}
\usepackage{amsmath}
\usepackage{amssymb}
\usepackage{geometry}

\newtheorem{theorem}{Theorem}

\newtheorem{proposition}[theorem]{Proposition}

\newenvironment{proof}{\noindent\textbf{Proof:} }{\indent\hfill\rule{0.5em}{0.5em}}
\input{tcilatex}

\begin{document}

\title{Spinor fields without Lorentz frames in curved spacetime using
complexified quaternions}
\author{John Fredsted\thanks{%
physics@johnfredsted.dk} \\
%EndAName
Soeskraenten 22, Stavtrup, DK-8260 Viby J., Denmark}
\maketitle

\begin{abstract}
Using complexified quaternions, a formalism without Lorentz frames, and
therefore also without vierbeins, for dealing with tensor and spinor fields
in curved spacetime is presented. A local $\mathrm{U}\left( 1\right) $ gauge
symmetry, which, it is speculated, might be related to electromagnetism,
emerges naturally.
\end{abstract}

\section{Introduction}

Originating from the papers by Fock and Ivanenko \cite{Fock and Ivanenko},
and Weyl \cite{Weyl}, the standard formulation \cite{Weinberg
Gravitation,Weinberg Quantum,GSW} of spinor fields in curved spacetime
employs a vierbein field $e^{a}{}_{\mu }$ which erects at each point in
spacetime a local Lorentz frame. Using complexified quaternions, this paper
will present a coordinate covariant and locally Lorentz invariant formalism
for spinor fields in curved spacetime using no local Lorentz frames.

In general relativity which is our current best theory of the dynamics of
curved spacetime, the fundamental structures are the basis $\mathbf{e}_{\mu
}\in TM$, which may or may not be a coordinate basis $\partial _{\mu }$, the
dual $1$-form basis $\mathbf{\omega }^{\mu }\in T^{\ast }M$ obeying $\delta
_{\nu }^{\mu }\equiv \left\langle \mathbf{\omega }^{\mu },\mathbf{e}_{\nu
}\right\rangle $, the metric $\mathbf{g}$ with components $g_{\mu \nu
}\equiv \mathbf{g}\left( \mathbf{e}_{\mu }\mathbf{,e}_{\nu }\right) \equiv 
\mathbf{e}_{\mu }\cdot \mathbf{e}_{\nu }$, and the connection coefficients $%
\Gamma ^{\mu }{}_{\nu \rho }\equiv \left\langle \mathbf{\omega }^{\mu
},\nabla _{\rho }\mathbf{e}_{\nu }\right\rangle $ with $\nabla _{\rho }$ the
covariant derivative \cite{MTW}. For dealing with spinor fields, this
structure is augmented by a vierbein field $e^{a}{}_{\mu }$, as previously
mentioned, obeying $g_{\mu \nu }=\allowbreak \eta _{ab}e^{a}{}_{\mu
}e^{b}{}_{\nu }$, where $\eta _{ab}$ is the Minkowski metric, and a spin
connection $\omega _{\mu }{}^{ab}$ which in its minimal version is given by $%
\omega _{\mu }{}^{ab}\equiv g^{\rho \sigma }e^{a}{}_{\rho }\nabla _{\mu
}e^{b}{}_{\sigma }$ \cite{Weinberg Quantum}.

This paper will present a formalism able of dealing with both tensor and
spinor fields, using much less structure (see the next section for the
notation): a basis $\mathrm{s}_{\mu }\in \left( \mathbb{C}\otimes \mathbb{H}%
\right) ^{-}$ and a spin connection $\omega _{\mu }\in \mathbb{C}\otimes 
\mathbb{H}$ obeying $0=\mathrm{Scal}\left( \omega _{\mu }+\overline{\omega }%
_{\mu }^{\ast }\right) $, Eq. (\ref{Eq:OmegaZeroScal}), in terms of which
both the metric and the 'minimal version', to be specified later, of the
connection coefficients are determined as $g_{\mu \nu }\equiv \left\langle 
\mathrm{s}_{\mu },\mathrm{s}_{\nu }\right\rangle $, Eq. (\ref{Eq:Metric}),
and $\Gamma ^{\mu }{}_{\nu \rho }\equiv \left\langle \mathrm{s}^{\mu
},\partial _{\rho }\mathrm{s}_{\nu }+\omega _{\rho }\mathrm{s}_{\nu }+%
\mathrm{s}_{\nu }\overline{\omega }_{\rho }^{\ast }\right\rangle $, Eq. (\ref%
{Eq:MinimalConn}), respectively.

The condition $0=\mathrm{Scal}\left( \omega _{\mu }+\overline{\omega }_{\mu
}^{\ast }\right) $ is equivalent to $\omega _{\mu }\in \mathrm{i}\mathbb{R}%
\vee \omega _{\mu }\in \mathbb{C}\otimes \mathrm{Vec}\left( \mathbb{H}%
\right) $. The freedom $\omega _{\mu }\in \mathrm{i}\mathbb{R}$ is related
to a local $\mathrm{U}\left( 1\right) $ gauge freedom, Eq. (\ref{Eq:OmegaU1}%
), which, it is speculated, might be related to electromagnetism.

\section{Notation}

The set of complexified quaternions is denoted $\mathbb{C}\otimes \mathbb{H}$%
, which is equal to $\mathbb{H}\otimes \mathbb{C}$ because $\mathbb{C}$ and $%
\mathbb{H}$ are assumed to commute. Usual complex conjugation acts as $%
\mathbb{C}\otimes \mathbb{H\rightarrow C}^{\ast }\otimes \mathbb{H}$, and
usual quaternionic conjugation \cite{Okubo,Springer and Veldkamp} acts as $%
\mathbb{C}\otimes \mathbb{H\rightarrow C}\otimes \overline{\mathbb{H}}$. In
conjunction, these two conjugations may be used to split the complexified
quaternions as $\mathbb{C}\otimes \mathbb{H}=\left( \mathbb{C}\otimes 
\mathbb{H}\right) ^{-}\cup \left( \mathbb{C}\otimes \mathbb{H}\right) ^{+}$,
where%
\begin{equation*}
\left( \mathbb{C}\otimes \mathbb{H}\right) ^{\pm }\equiv \left\{ x\in 
\mathbb{C}\otimes \mathbb{H}\left\vert \overline{x}^{\ast }=\pm x\right.
\right\} .
\end{equation*}%
The real- and imaginary parts of $\mathbb{C}$ are, as usual, denoted $\func{%
Re}\left( \mathbb{C}\right) =\mathbb{R}$ and $\func{Im}\left( \mathbb{C}%
\right) =\mathrm{i}\mathbb{R}$, respectively. The scalar- and vector parts
of $\mathbb{H}$, denoted $\mathrm{Scal}\left( \mathbb{H}\right) $ and $%
\mathrm{Vec}\left( \mathbb{H}\right) $, respectively, are defined as%
\begin{eqnarray*}
\mathrm{Scal}\left( \mathbb{H}\right) &\equiv &\left\{ x\in \mathbb{H}%
\left\vert \overline{x}=+x\right. \right\} , \\
\mathrm{Vec}\left( \mathbb{H}\right) &\equiv &\left\{ x\in \mathbb{H}%
\left\vert \overline{x}=-x\right. \right\} .
\end{eqnarray*}%
The bilinear inner product $\left\langle \cdot ,\cdot \right\rangle :\left( 
\mathbb{C}\otimes \mathbb{H}\right) ^{2}\mathbb{\rightarrow C}$ is defined
as \cite{Dundarer and Gursey}%
\begin{equation}
2\left\langle x,y\right\rangle \equiv x\overline{y}+y\overline{x}\equiv 
\overline{x}y+\overline{y}x.  \label{Eq:ipDef}
\end{equation}

\section{Basis and metric}

Let $\left( M,x^{\mu }\right) $ be some spacetime four-manifold $M$
parametrized by some coordinate system $x^{\mu }$. In order to endow $M$
with a measure of distance, a metric $g_{\mu \nu }\in \mathbb{R}$ must be
provided. For $M$ a Riemannian manifold, the metric is given by $g_{\mu \nu
}\equiv \mathbf{g}\left( \mathbf{e}_{\mu }\mathbf{,e}_{\nu }\right) $ with $%
\mathbf{g}\equiv g_{\mu \nu }\mathbf{\omega }^{\mu }\otimes \mathbf{\omega }%
^{\nu }$, where $\mathbf{e}_{\mu }\in TM$ is the basis, and $\mathbf{\omega }%
^{\mu }\in T^{\ast }M$ is the dual $1$-form basis, obeying $\delta _{\nu
}^{\mu }=\left\langle \mathbf{\omega }^{\mu },\mathbf{e}_{\nu }\right\rangle 
$ \cite{MTW}.

This route will not be taken in this paper: neither a basis $\mathbf{e}_{\mu
}\in TM$ nor a dual $1$-form basis $\mathbf{\omega }^{\mu }\in T^{\ast }M$
will be introduced, thus leaving the arena of Riemannian manifolds. Instead
the metric $g_{\mu \nu }$ will be given by%
\begin{eqnarray}
\mathbb{R}\ni 2g_{\mu \nu } &\equiv &2\left\langle \mathrm{s}_{\mu },\mathrm{%
s}_{\nu }\right\rangle  \label{Eq:Metric} \\
&\equiv &\mathrm{s}_{\mu }\overline{\mathrm{s}}_{\nu }+\mathrm{s}_{\nu }%
\overline{\mathrm{s}}_{\mu }  \notag \\
&\equiv &\overline{\mathrm{s}}_{\mu }\mathrm{s}_{\nu }+\overline{\mathrm{s}}%
_{\nu }\mathrm{s}_{\mu },  \notag
\end{eqnarray}%
with $\mathrm{s}_{\mu }=-\overline{\mathrm{s}}_{\mu }^{\ast }\in \left( 
\mathbb{C}\otimes \mathbb{H}\right) ^{-}$ the basis obeying the
nondegeneracy condition%
\begin{equation}
\left\{ c^{\mu }\mathrm{s}_{\mu }\left\vert c^{\mu }\in \mathbb{C}\right.
\right\} =\mathbb{C}\otimes \mathbb{H}.  \label{Eq:Nondegeneracy}
\end{equation}%
A basis $\mathrm{s}_{\mu }=+\overline{\mathrm{s}}_{\mu }^{\ast }\in \left( 
\mathbb{C}\otimes \mathbb{H}\right) ^{+}$, corresponding to the opposite
signature of the metric, could, of course, equally well have been chosen.
The real-valuedness of the metric, as indicated above, is a direct
consequence of Proposition \ref{Prop:RealValuedIP}. The symmetry of the
metric follows trivially from the symmetry of the inner product, $%
\left\langle x,y\right\rangle \equiv \left\langle y,x\right\rangle $.

The nondegeneracy property of $\mathrm{s}_{\mu }$, as stated above, implies
the nondegeneracy of the metric, i.e., $\det \left( g_{\mu \nu }\right) \neq
0$, as follows: Assume that $\left\{ c^{\mu }\mathrm{s}_{\mu }\left\vert
c^{\mu }\in \mathbb{C}\right. \right\} =\mathbb{C}\otimes \mathbb{H}$, but $%
\det \left( g_{\mu \nu }\right) =0$. Then, performing matrix row operations,
there must exist some linear combination $d^{\mu }\mathrm{s}_{\mu }$, with $%
d^{\mu }\in \mathbb{C}$, so that $\left\langle d^{\mu }\mathrm{s}_{\mu },%
\mathrm{s}_{\nu }\right\rangle =0$ for all $\mathrm{s}_{\nu }$, which
implies that $d^{\mu }\mathrm{s}_{\mu }=0$, which implies that $\dim \left(
\left\{ c^{\mu }\mathrm{s}_{\mu }\left\vert c^{\mu }\in \mathbb{C}\right.
\right\} \right) <\dim \left( \mathbb{C}\otimes \mathbb{H}\right) $. A
contradiction.

The nondegeneracy of the metric allows the definition $\mathrm{s}^{\mu
}\equiv g^{\mu \nu }\mathrm{s}_{\nu }\Leftrightarrow \mathrm{s}_{\mu
}=g_{\mu \nu }\mathrm{s}^{\nu }$, which, unsurprisingly, obeys $\delta _{\nu
}^{\mu }=\left\langle \mathrm{s}^{\mu },\mathrm{s}_{\nu }\right\rangle $ and 
$g^{\mu \nu }=\left\langle \mathrm{s}^{\mu },\mathrm{s}^{\nu }\right\rangle $%
. Thus, as usual, coordinate indices are raised and lowered with the inverse
metric and the metric, respectively. Note that $\delta _{\nu }^{\mu
}=\left\langle \mathrm{s}^{\mu },\mathrm{s}_{\nu }\right\rangle $ may be
considered the analogue of $\delta _{\nu }^{\mu }=\left\langle \mathbf{%
\omega }^{\mu },\mathbf{e}_{\nu }\right\rangle $ in Riemannian calculus, but
whereas it is quite sensible to raise and lower indices in $\left\langle 
\mathrm{s}^{\mu },\mathrm{s}_{\nu }\right\rangle $, it is not in $%
\left\langle \mathbf{\omega }^{\mu },\mathbf{e}_{\nu }\right\rangle $,
because $\mathbf{e}_{\mu }$ and $\mathbf{\omega }^{\mu }$ are \textit{fixed}
as type $\left( 0,1\right) $ and $\left( 1,0\right) $ tensors, respectively.

The basis $\mathrm{s}_{\mu }$ is subject to two different types of
transformations:

\begin{itemize}
\item Coordinate transformations: In accordance with its coordinate index $%
\mu $, under a coordinate transformation $dx^{\prime \mu }=\left( \partial
x^{\prime \mu }/\partial x^{\nu }\right) dx^{\nu }$, the basis $\mathrm{s}%
_{\mu }$ is assumed to transform as a type $\left( 0,1\right) $ tensor, i.e.,%
\begin{equation}
\mathrm{s}_{\mu }^{\prime }=\frac{\partial x^{\nu }}{\partial x^{\prime \mu }%
}\mathrm{s}_{\nu },  \label{Eq:CoordTransfBasis}
\end{equation}%
thus making $g_{\mu \nu }$ and $\mathrm{s}^{\mu }$ transform as type $\left(
0,2\right) $ and type $\left( 1,0\right) $ tensors, respectively.

\item Local Lorentz transformations: Under the local transformation $\mathrm{%
s}_{\mu }^{\prime }=\Lambda \mathrm{s}_{\mu }\overline{\Lambda }^{\ast }$,
where $\Lambda \equiv \Lambda \left( x^{\mu }\right) $ obeys $\overline{%
\Lambda }=\Lambda ^{-1}\Leftrightarrow 1=\Lambda \overline{\Lambda }=%
\overline{\Lambda }\Lambda $, the metric is invariant as the following
calculation, using some of the identities of Proposition \ref%
{Prop:CompAlgIds}, shows:%
\begin{eqnarray}
g_{\mu \nu }^{\prime } &=&\left\langle \Lambda \mathrm{s}_{\mu }\overline{%
\Lambda }^{\ast },\Lambda \mathrm{s}_{\nu }\overline{\Lambda }^{\ast
}\right\rangle  \notag \\
&=&\left\langle \left( \overline{\Lambda }\Lambda \right) \mathrm{s}_{\mu },%
\mathrm{s}_{\nu }\left( \overline{\Lambda }\Lambda \right) ^{\ast
}\right\rangle  \notag \\
&=&g_{\mu \nu }.  \label{Eq:TransfMetric}
\end{eqnarray}%
Note that $\mathrm{s}_{\mu }\rightarrow \mathrm{s}_{\mu }^{\prime }=\Lambda 
\mathrm{s}_{\mu }\overline{\Lambda }^{\ast }$ maps, as it should, from $%
\left( \mathbb{C}\otimes \mathbb{H}\right) ^{-}$ to $\left( \mathbb{C}%
\otimes \mathbb{H}\right) ^{-}$. The associativity of the (complexified)
quaternions makes it unnecessary to care about parantheses in the
calculation. The group%
\begin{equation*}
\left\langle \left\{ \Lambda \in \mathbb{C}\otimes \mathbb{H}\left\vert 
\overline{\Lambda }=\Lambda ^{-1}\right. \right\} ,\times ,1\right\rangle
\end{equation*}%
with multiplication $\times $ as group operation, and $1$ as identity
element is isomorphic to $\mathrm{Sp}\left( 1,\mathbb{C}\right) \cong 
\mathrm{SL}\left( 2,\mathbb{C}\right) $, the double cover of the Lorentz
group $\mathrm{SO}\left( 3,1\right) $. Therefore, the transformation $%
\mathrm{s}_{\mu }^{\prime }=\Lambda \mathrm{s}_{\mu }\overline{\Lambda }%
^{\ast }$ is a proper Lorentz transformation \cite{Lambek}.
\end{itemize}

\section{Covariant derivative and connections}

Under the local Lorentz transformation $\mathrm{s}_{\mu }^{\prime }=\Lambda 
\mathrm{s}_{\mu }\overline{\Lambda }^{\ast }$, the derivative $\partial
_{\mu }\mathrm{s}_{\nu }$ transforms noncovariantly, i.e., $\left( \partial
_{\mu }\mathrm{s}_{\nu }\right) ^{\prime }\neq \Lambda \left( \partial _{\mu
}\mathrm{s}_{\nu }\right) \overline{\Lambda }^{\ast }$, generally. The
remedy, as usual, is to define a covariant derivative $D_{\mu }$ for which $%
\left( D_{\mu }\mathrm{s}_{\nu }\right) ^{\prime }=\Lambda \left( D_{\mu }%
\mathrm{s}_{\nu }\right) \overline{\Lambda }^{\ast }$. Preferably, $D_{\mu }$
should commute with both complex- and quaternionic conjugation, i.e.,%
\begin{eqnarray}
D_{\mu }\phi ^{\ast } &\equiv &\left( D_{\mu }\phi \right) ^{\ast },
\label{Eq:CovDerComplConj} \\
D_{\mu }\overline{\phi } &\equiv &\overline{\left( D_{\mu }\phi \right) },
\label{Eq:CovDerQuatConj}
\end{eqnarray}%
for all fields $\phi \in \mathbb{C}\otimes \mathbb{H}$ (any coordinate
indices suppressed). Before defining the covariant derivative of $\mathrm{s}%
_{\mu }$, the socalled quaternionic spinor fields, quaternionic vector
field, and quaternionic scalar field, and their covariant derivatives, will
be defined.

\subsection{Quaternionic spinor fields}

Fields $\psi _{L},\psi _{R}\in \mathbb{C}\otimes \mathbb{H}$ that transform
as $\psi _{L}^{\prime }=\Lambda \psi _{L}$ and $\psi _{R}^{\prime }=\psi _{R}%
\overline{\Lambda }^{\ast }$, respectively, will be called \textit{%
quaternionic spinor fields}. Their covariant derivatives are defined as%
\begin{eqnarray}
D_{\mu }\psi _{L} &\equiv &\partial _{\mu }\psi _{L}+\omega _{\mu }\psi _{L},
\label{Eq:CovDerPsiL} \\
D_{\mu }\psi _{R} &\equiv &\partial _{\mu }\psi _{R}+\psi _{R}\overline{%
\omega }_{\mu }^{\ast },  \label{Eq:CovDerPsiR}
\end{eqnarray}%
where $\omega _{\mu }\in \mathbb{C}\otimes \mathbb{H}$ is some gauge
connection obeying $0=\mathrm{Scal}\left( \omega _{\mu }+\overline{\omega }%
_{\mu }^{\ast }\right) $, a condition which will become clear shortly.

\begin{proposition}
\label{Prop:CovDerSpinors}The covariant derivatives $D_{\mu }\psi _{L}$ and $%
D_{\mu }\psi _{R}$ transform covariantly,%
\begin{eqnarray}
\left( D_{\mu }\psi _{L}\right) ^{\prime } &=&\Lambda \left( D_{\mu }\psi
_{L}\right) ,  \label{Eq:TransfPsiL} \\
\left( D_{\mu }\psi _{R}\right) ^{\prime } &=&\left( D_{\mu }\psi
_{R}\right) \overline{\Lambda }^{\ast },  \label{Eq:TransfPsiR}
\end{eqnarray}%
if and only if the gauge connection $\omega _{\mu }$ transforms as%
\begin{equation}
\omega _{\mu }^{\prime }=\Lambda \omega _{\mu }\overline{\Lambda }-\left(
\partial _{\mu }\Lambda \right) \overline{\Lambda }.
\label{Eq:TransfLorentzConn}
\end{equation}
\end{proposition}

\begin{proof}
By direct calculation follows%
\begin{eqnarray*}
\left( D_{\mu }\psi _{L}\right) ^{\prime } &=&\left( \partial _{\mu }\Lambda
\right) \psi _{L}+\Lambda \left( \partial _{\mu }\psi _{L}\right) +\omega
_{\mu }^{\prime }\left( \Lambda \psi _{L}\right) , \\
\left( D_{\mu }\psi _{R}\right) ^{\prime } &=&\left( \partial _{\mu }\psi
_{R}\right) \overline{\Lambda }^{\ast }+\psi _{R}\left( \partial _{\mu }%
\overline{\Lambda }^{\ast }\right) +\left( \psi _{R}\overline{\Lambda }%
^{\ast }\right) \overline{\left( \omega _{\mu }^{\prime }\right) }^{\ast },
\end{eqnarray*}%
which equal $\Lambda \left( D_{\mu }\psi _{L}\right) $ and $\left( D_{\mu
}\psi _{R}\right) \overline{\Lambda }^{\ast }$, respectively, if and only if%
\begin{eqnarray*}
\left( \partial _{\mu }\Lambda \right) \psi _{L}+\omega _{\mu }^{\prime
}\left( \Lambda \psi _{L}\right) &=&\Lambda \left( \omega _{\mu }\psi
_{L}\right) , \\
\psi _{R}\left( \partial _{\mu }\overline{\Lambda }^{\ast }\right) +\left(
\psi _{R}\overline{\Lambda }^{\ast }\right) \overline{\left( \omega _{\mu
}^{\prime }\right) }^{\ast } &=&\left( \psi _{R}\overline{\omega }_{\mu
}^{\ast }\right) \overline{\Lambda }^{\ast },
\end{eqnarray*}%
which, using the associativity of the complexified quaternions to move
parantheses, the arbitrariness of $\psi _{L}$ and $\psi _{R}$, and $%
\overline{\Lambda }=\Lambda ^{-1}$, imply the two equivalent conditions%
\begin{eqnarray*}
\omega _{\mu }^{\prime } &=&\Lambda \omega _{\mu }\overline{\Lambda }-\left(
\partial _{\mu }\Lambda \right) \overline{\Lambda }, \\
\overline{\left( \omega _{\mu }^{\prime }\right) }^{\ast } &=&\Lambda ^{\ast
}\overline{\omega }_{\mu }^{\ast }\overline{\Lambda }^{\ast }-\Lambda ^{\ast
}\left( \partial _{\mu }\overline{\Lambda }^{\ast }\right) .
\end{eqnarray*}
\end{proof}

Generally, fields $L_{\sigma _{1}\cdots \sigma _{n}}^{\rho _{1}\cdots \rho
_{m}},R_{\sigma _{1}\cdots \sigma _{n}}^{\rho _{1}\cdots \rho _{m}}\in 
\mathbb{C}\otimes \mathbb{H}$ that transform as $\left( L_{\sigma _{1}\cdots
\sigma _{n}}^{\rho _{1}\cdots \rho _{m}}\right) ^{\prime }=\Lambda L_{\sigma
_{1}\cdots \sigma _{n}}^{\rho _{1}\cdots \rho _{m}}$ and $\left( R_{\sigma
_{1}\cdots \sigma _{n}}^{\rho _{1}\cdots \rho _{m}}\right) ^{\prime
}=R_{\sigma _{1}\cdots \sigma _{n}}^{\rho _{1}\cdots \rho _{m}}\overline{%
\Lambda }^{\ast }$, respectively, will be called \textit{type }$\left(
m,n\right) $\textit{\ tensor-valued quaternionic spinor fields}. The
covariant derivative of these fields are defined as%
\begin{eqnarray}
D_{\mu }L_{\sigma _{1}\cdots \sigma _{n}}^{\rho _{1}\cdots \rho _{m}}
&\equiv &\nabla _{\mu }L_{\sigma _{1}\cdots \sigma _{n}}^{\rho _{1}\cdots
\rho _{m}}+\omega _{\mu }L_{\sigma _{1}\cdots \sigma _{n}}^{\rho _{1}\cdots
\rho _{m}},  \label{Eq:CovDerPsiLTensor} \\
D_{\mu }R_{\sigma _{1}\cdots \sigma _{n}}^{\rho _{1}\cdots \rho _{m}}
&\equiv &\nabla _{\mu }R_{\sigma _{1}\cdots \sigma _{n}}^{\rho _{1}\cdots
\rho _{m}}+R_{\sigma _{1}\cdots \sigma _{n}}^{\rho _{1}\cdots \rho _{m}}%
\overline{\omega }_{\mu }^{\ast }.  \label{Eq:CovDerPsiRTensor}
\end{eqnarray}%
In comparison with $D_{\mu }\psi _{L}$ and $D_{\mu }\psi _{R}$, above, note
here the explicit appearance of $\nabla _{\mu }$, which is necessary because
the quaternionic spinor fields $L_{\sigma _{1}\cdots \sigma _{n}}^{\rho
_{1}\cdots \rho _{m}}$ and $R_{\sigma _{1}\cdots \sigma _{n}}^{\rho
_{1}\cdots \rho _{m}}$ now carry coordinate indices. They transform
covariantly;%
\begin{eqnarray}
\left( D_{\mu }L_{\sigma _{1}\cdots \sigma _{n}}^{\rho _{1}\cdots \rho
_{m}}\right) ^{\prime } &=&\Lambda \left( D_{\mu }L_{\sigma _{1}\cdots
\sigma _{n}}^{\rho _{1}\cdots \rho _{m}}\right) ,
\label{Eq:TransfPsiLTensor} \\
\left( D_{\mu }R_{\sigma _{1}\cdots \sigma _{n}}^{\rho _{1}\cdots \rho
_{m}}\right) ^{\prime } &=&\left( D_{\mu }R_{\sigma _{1}\cdots \sigma
_{n}}^{\rho _{1}\cdots \rho _{m}}\right) \overline{\Lambda }^{\ast },
\label{Eq:TransfPsiRTensor}
\end{eqnarray}%
the proof of which is completely analogous to the proof above for $D_{\mu
}\psi _{L}$ and $D_{\mu }\psi _{R}$. The above fields $\psi _{L}$ and $\psi
_{R}$ are, of course, type $\left( 0,0\right) $ tensor-valued quaternionic
spinor fields.

\subsection{Quaternionic vector field}

A field $V\in \mathbb{C}\otimes \mathbb{H}$ that transforms as $V^{\prime
}=\Lambda V\overline{\Lambda }^{\ast }$ will be called a \textit{%
quaternionic vector field}. Its covariant derivative is defined as%
\begin{equation}
D_{\mu }V\equiv \partial _{\mu }V+\omega _{\mu }V+V\overline{\omega }_{\mu
}^{\ast }.  \label{Eq:CovDerVector}
\end{equation}

\begin{proposition}
\label{Prop:CovDerVector}The covariant derivative $D_{\mu }V$ transforms
covariantly,%
\begin{equation}
\left( D_{\mu }V\right) ^{\prime }=\Lambda \left( D_{\mu }V\right) \overline{%
\Lambda }^{\ast }.  \label{Eq:TransfVector}
\end{equation}
\end{proposition}

\begin{proof}
Follows by direct calculation:%
\begin{eqnarray*}
\left( D_{\mu }V\right) ^{\prime } &\equiv &\partial _{\mu }V^{\prime
}+\omega _{\mu }^{\prime }V^{\prime }+V^{\prime }\overline{\left( \omega
_{\mu }^{\prime }\right) }^{\ast } \\
&=&\partial _{\mu }\left( \Lambda V\overline{\Lambda }^{\ast }\right) \\
&&+\left[ \Lambda \omega _{\mu }\overline{\Lambda }-\left( \partial _{\mu
}\Lambda \right) \overline{\Lambda }\right] \left( \Lambda V\overline{%
\Lambda }^{\ast }\right) \\
&&+\left( \Lambda V\overline{\Lambda }^{\ast }\right) \left[ \Lambda ^{\ast }%
\overline{\omega }_{\mu }^{\ast }\overline{\Lambda }^{\ast }-\Lambda ^{\ast
}\left( \partial _{\mu }\overline{\Lambda }^{\ast }\right) \right] \\
&=&\left( \partial _{\mu }\Lambda \right) V\overline{\Lambda }^{\ast
}+\Lambda \left( \partial _{\mu }V\right) \overline{\Lambda }^{\ast
}+\Lambda V\left( \partial _{\mu }\overline{\Lambda }^{\ast }\right) \\
&&+\Lambda \omega _{\mu }\left( \overline{\Lambda }\Lambda \right) V%
\overline{\Lambda }^{\ast }-\left( \partial _{\mu }\Lambda \right) \left( 
\overline{\Lambda }\Lambda \right) V\overline{\Lambda }^{\ast } \\
&&+\Lambda V\left( \overline{\Lambda }^{\ast }\Lambda ^{\ast }\right) 
\overline{\omega }_{\mu }^{\ast }\overline{\Lambda }^{\ast }-\Lambda V\left( 
\overline{\Lambda }^{\ast }\Lambda ^{\ast }\right) \left( \partial _{\mu }%
\overline{\Lambda }^{\ast }\right) ,
\end{eqnarray*}%
which, using $1=\overline{\Lambda }\Lambda =\overline{\Lambda }^{\ast
}\Lambda ^{\ast }$, implies%
\begin{equation*}
\left( D_{\mu }V\right) ^{\prime }=\Lambda \left( \partial _{\mu }V+\omega
_{\mu }V+V\overline{\omega }_{\mu }^{\ast }\right) \overline{\Lambda }^{\ast
}=\Lambda \left( D_{\mu }V\right) \overline{\Lambda }^{\ast }.
\end{equation*}
\end{proof}

Generally, a field $V_{\sigma _{1}\cdots \sigma _{n}}^{\rho _{1}\cdots \rho
_{m}}\in \mathbb{C}\otimes \mathbb{H}$ that transform as $\left( V_{\sigma
_{1}\cdots \sigma _{n}}^{\rho _{1}\cdots \rho _{m}}\right) ^{\prime
}=\Lambda V_{\sigma _{1}\cdots \sigma _{n}}^{\rho _{1}\cdots \rho _{m}}%
\overline{\Lambda }^{\ast }$ will be called a \textit{type }$\left(
m,n\right) $\textit{\ tensor-valued quaternionic vector field}. The
covariant derivative of this field is defined as%
\begin{equation}
D_{\mu }V_{\sigma _{1}\cdots \sigma _{n}}^{\rho _{1}\cdots \rho _{m}}=\nabla
_{\mu }V_{\sigma _{1}\cdots \sigma _{n}}^{\rho _{1}\cdots \rho _{m}}+\omega
_{\mu }V_{\sigma _{1}\cdots \sigma _{n}}^{\rho _{1}\cdots \rho
_{m}}+V_{\sigma _{1}\cdots \sigma _{n}}^{\rho _{1}\cdots \rho _{m}}\overline{%
\omega }_{\mu }^{\ast }.  \label{Eq:CovDerVectorTensor}
\end{equation}%
Analogous to $D_{\mu }L_{\sigma _{1}\cdots \sigma _{n}}^{\rho _{1}\cdots
\rho _{m}}$ and $D_{\mu }R_{\sigma _{1}\cdots \sigma _{n}}^{\rho _{1}\cdots
\rho _{m}}$, above, $\nabla _{\mu }$ is necessary here because the
quaternionic vector field $V_{\sigma _{1}\cdots \sigma _{n}}^{\rho
_{1}\cdots \rho _{m}}$ now carry coordinate indices. It transforms
covariantly;%
\begin{equation}
\left( D_{\mu }V_{\sigma _{1}\cdots \sigma _{n}}^{\rho _{1}\cdots \rho
_{m}}\right) ^{\prime }=\Lambda \left( D_{\mu }V_{\sigma _{1}\cdots \sigma
_{n}}^{\rho _{1}\cdots \rho _{m}}\right) \overline{\Lambda }^{\ast },
\label{Eq:TransfVectorTensor}
\end{equation}%
the proof of which is completely analogous to the proof above for $D_{\mu }V$%
. The above field $V$ is, of course, a type $\left( 0,0\right) $
tensor-valued quaternionic vector field.

\subsection{Quaternionic scalar field}

A field $S\in \mathbb{C}\otimes \mathbb{H}$ that transforms invariantly, $%
S^{\prime }=S$, will be called a \textit{quaternionic scalar field}. Its
covariant derivative is defined as%
\begin{equation}
D_{\mu }S\equiv \partial _{\mu }S.  \label{Eq:CovDerScalar}
\end{equation}

Generally, a field $S_{\sigma _{1}\cdots \sigma _{n}}^{\rho _{1}\cdots \rho
_{m}}\in \mathbb{C}\otimes \mathbb{H}$ that transform as $\left( S_{\sigma
_{1}\cdots \sigma _{n}}^{\rho _{1}\cdots \rho _{m}}\right) ^{\prime
}=S_{\sigma _{1}\cdots \sigma _{n}}^{\rho _{1}\cdots \rho _{m}}$ will be
called a \textit{type }$\left( m,n\right) $\textit{\ tensor-valued
quaternionic scalar field}. The covariant derivative of this field is
defined as%
\begin{equation}
D_{\mu }S_{\sigma _{1}\cdots \sigma _{n}}^{\rho _{1}\cdots \rho _{m}}=\nabla
_{\mu }S_{\sigma _{1}\cdots \sigma _{n}}^{\rho _{1}\cdots \rho _{m}}.
\label{Eq:CovDerScalarTensor}
\end{equation}%
Analogous to $D_{\mu }V_{\sigma _{1}\cdots \sigma _{n}}^{\rho _{1}\cdots
\rho _{m}}$, above, $\nabla _{\mu }$ is necessary here because the
quaternionic scalar field $S_{\sigma _{1}\cdots \sigma _{n}}^{\rho
_{1}\cdots \rho _{m}}$ now carry coordinate indices. It transforms
covariantly (and invariantly);%
\begin{equation}
\left( D_{\mu }S_{\sigma _{1}\cdots \sigma _{n}}^{\rho _{1}\cdots \rho
_{m}}\right) ^{\prime }=D_{\mu }S_{\sigma _{1}\cdots \sigma _{n}}^{\rho
_{1}\cdots \rho _{m}}.  \label{Eq:TransfScalarTensor}
\end{equation}%
The above field $S$ is, of course, a type $\left( 0,0\right) $ tensor-valued
quaternionic scalar field.

\subsection{Basis and metric}

The basis field $\mathrm{s}_{\mu }$ is a type $\left( 0,1\right) $
tensor-valued quaternionic vector field, so according to Eq. (\ref%
{Eq:CovDerVectorTensor}) its covariant derivative is defined as%
\begin{equation}
D_{\rho }\mathrm{s}_{\nu }\equiv \nabla _{\rho }\mathrm{s}_{\nu }+\omega
_{\rho }\mathrm{s}_{\nu }+\mathrm{s}_{\nu }\overline{\omega }_{\rho }^{\ast
}.  \label{Eq:CovDerBasis}
\end{equation}%
In complete analogy with how $\nabla _{\rho }$ acts in Riemannian calculus,
the covariant derivative $\nabla _{\rho }\mathrm{s}_{\nu }$ is defined as%
\begin{equation}
\nabla _{\rho }\mathrm{s}_{\nu }\equiv \partial _{\rho }\mathrm{s}_{\nu
}-\Gamma {}^{\mu }{}_{\nu \rho }\mathrm{s}_{\mu },
\label{Eq:NablaCovDerBasis}
\end{equation}%
where $\Gamma {}^{\mu }{}_{\nu \rho }$ are connection coefficients which
couple, in the usual way, to any coordinate index, covariant or
contravariant. Note, however, that a priori $\Gamma {}^{\mu }{}_{\nu \rho }$
have nothing to do with the usual connection coefficients $\Gamma ^{\mu
}{}_{\nu \rho }\equiv \left\langle \mathbf{\omega }^{\mu },\nabla _{\rho }%
\mathbf{e}_{\nu }\right\rangle $ of Riemannian calculus, because, as
previously stated, neither the basis $\mathbf{e}_{\mu }\in TM$ nor the dual $%
1$-form basis $\mathbf{\omega }^{\mu }\in T^{\ast }M$ is provided in the
present formalism.

So, as of yet, $\Gamma {}^{\mu }{}_{\nu \rho }$ are undetermined. A 'minimal
version' of $\Gamma {}^{\mu }{}_{\nu \rho }$, in the spirit of a minimal
spin connection \cite{Weinberg Quantum,GSW}, can be determined by solving
the 'minimality condition' $0\equiv D_{\rho }\mathrm{s}_{\nu }$ for $\Gamma
{}^{\mu }{}_{\nu \rho }=\Gamma {}^{\mu }{}_{\nu \rho }\left( \mathrm{s}%
_{\sigma },\omega _{\sigma }\right) $, the result being%
\begin{equation}
\mathbb{R}\ni \Gamma {}^{\mu }{}_{\nu \rho }\equiv \left\langle \mathrm{s}%
^{\mu },\partial _{\rho }\mathrm{s}_{\nu }+\omega _{\rho }\mathrm{s}_{\nu }+%
\mathrm{s}_{\nu }\overline{\omega }_{\rho }^{\ast }\right\rangle .
\label{Eq:MinimalConn}
\end{equation}

Expressing $\Gamma {}^{\mu }{}_{\nu \rho }$ in terms of $\mathrm{s}_{\mu }$
and $\omega _{\mu }$, which is somehow the reverse of expressing a minimal
spin connection $\omega _{\mu }{}^{ab}$ in terms of a vierbein $e^{a}{}_{\mu
}$ and the connection coefficients $\Gamma {}^{\mu }{}_{\nu \rho }$, as is
customarily done \cite{Weinberg Quantum,GSW}, is reasonable because with
neither $\mathbf{e}_{\mu }\in TM$ nor $\mathbf{\omega }^{\mu }\in T^{\ast }M$
provided in the present formalism, the connection coefficients seem to 'flap
in the breeze': the only potentially reasonable way, it seems, that they
could be expressed without using $\omega _{\mu }$ is as the usual
Christoffel coefficients,%
\begin{equation*}
\left\{ _{\mu \nu }^{\rho }\right\} \equiv \frac{1}{2}g^{\rho \sigma }\left(
\partial _{\mu }g_{\sigma \nu }+\partial _{\nu }g_{\mu \sigma }-\partial
_{\sigma }g_{\mu \nu }\right) ,
\end{equation*}%
determined entirely in terms of the metric $g_{\mu \nu }=\left\langle 
\mathrm{s}_{\mu },\mathrm{s}_{\nu }\right\rangle $. But, remembering that
the Christoffel connection coefficients are the (general) connection
coefficients in a coordinate basis $\mathbf{e}_{\mu }=\partial _{\mu }$, why
should $\Gamma {}^{\rho }{}_{\mu \nu }$, as defined by Eq. (\ref%
{Eq:MinimalConn}), equal $\left\{ _{\mu \nu }^{\rho }\right\} $, when there
is no relation between the coordinate system $x^{\mu }$ and the basis $%
\mathrm{s}_{\mu }$?

Four important properties of $\Gamma {}^{\mu }{}_{\nu \rho }$, as defined by
Eq. (\ref{Eq:MinimalConn}), should be emphasized:

\begin{itemize}
\item Its real-valuedness, as indicated above: This follows directly from $%
\mathrm{s}^{\mu }\in \left( \mathbb{C}\otimes \mathbb{H}\right) ^{-}$ and $%
\partial _{\rho }\mathrm{s}_{\nu }+\omega _{\rho }\mathrm{s}_{\nu }+\mathrm{s%
}_{\nu }\overline{\omega }_{\rho }^{\ast }\in \left( \mathbb{C}\otimes 
\mathbb{H}\right) ^{-}$, using Proposition \ref{Prop:RealValuedIP}.

\item Its metric compatibility in the usual Riemannian sense, $0\equiv
\nabla _{\rho }g_{\mu \nu }$: Follows by direct calculation, using some of
the identities of Proposition \ref{Prop:CompAlgIds}:%
\begin{eqnarray*}
\nabla _{\rho }g_{\mu \nu } &\equiv &\partial _{\rho }g_{\mu \nu }-\Gamma
{}^{\sigma }{}_{\mu \rho }g_{\sigma \nu }-\Gamma {}^{\sigma }{}_{\nu \rho
}g_{\mu \sigma } \\
&=&\partial _{\rho }g_{\mu \nu }-\left\langle \mathrm{s}^{\sigma },\partial
_{\rho }\mathrm{s}_{\mu }+\omega _{\rho }\mathrm{s}_{\mu }+\mathrm{s}_{\mu }%
\overline{\omega }_{\rho }^{\ast }\right\rangle g_{\sigma \nu }-\left\langle 
\mathrm{s}^{\sigma },\partial _{\rho }\mathrm{s}_{\nu }+\omega _{\rho }%
\mathrm{s}_{\nu }+\mathrm{s}_{\nu }\overline{\omega }_{\rho }^{\ast
}\right\rangle g_{\mu \sigma } \\
&=&-\left\langle \mathrm{s}_{\nu },\omega _{\rho }\mathrm{s}_{\mu
}\right\rangle -\left\langle \mathrm{s}_{\nu },\mathrm{s}_{\mu }\overline{%
\omega }_{\rho }^{\ast }\right\rangle -\left\langle \mathrm{s}_{\mu },\omega
_{\rho }\mathrm{s}_{\nu }\right\rangle -\left\langle \mathrm{s}_{\mu },%
\mathrm{s}_{\nu }\overline{\omega }_{\rho }^{\ast }\right\rangle \\
&=&-\left\langle \mathrm{s}_{\nu }\overline{\mathrm{s}}_{\mu },\omega _{\rho
}\right\rangle -\left\langle \overline{\mathrm{s}}_{\mu }\mathrm{s}_{\nu },%
\overline{\omega }_{\rho }^{\ast }\right\rangle -\left\langle \mathrm{s}%
_{\mu }\overline{\mathrm{s}}_{\nu },\omega _{\rho }\right\rangle
-\left\langle \overline{\mathrm{s}}_{\nu }\mathrm{s}_{\mu },\overline{\omega 
}_{\rho }^{\ast }\right\rangle \\
&=&-\left\langle \mathrm{s}_{\mu }\overline{\mathrm{s}}_{\nu }+\mathrm{s}%
_{\nu }\overline{\mathrm{s}}_{\mu },\omega _{\rho }\right\rangle
-\left\langle \overline{\mathrm{s}}_{\mu }\mathrm{s}_{\nu }+\overline{%
\mathrm{s}}_{\nu }\mathrm{s}_{\mu },\overline{\omega }_{\rho }^{\ast
}\right\rangle \\
&=&-2g_{\mu \nu }\left\langle 1,\omega _{\rho }+\overline{\omega }_{\rho
}^{\ast }\right\rangle \\
&=&0,
\end{eqnarray*}%
where the last equality follows because%
\begin{equation}
0=\mathrm{Scal}\left( \omega _{\mu }+\overline{\omega }_{\mu }^{\ast
}\right) \Leftrightarrow \omega _{\mu }\in \mathrm{i}\mathbb{R}\vee \omega
_{\mu }\in \mathbb{C}\otimes \mathrm{Vec}\left( \mathbb{H}\right) ,
\label{Eq:OmegaZeroScal}
\end{equation}%
a condition on $\omega _{\mu }$ previously stated in connection with the
definitions of the covariant derivatives of the quaternionic spinor fields,
Eqs. (\ref{Eq:CovDerPsiL}) and (\ref{Eq:CovDerPsiR}).

\item Its transformation under coordinate transformations: By direct
calculation, using Eq. (\ref{Eq:MinimalConn}), and the fact that $\mathrm{s}%
_{\mu }$ and $\omega _{\mu }$ transform as type $\left( 0,1\right) $
tensors, it follows that $\Gamma {}^{\mu }{}_{\nu \rho }$ transform as the
connection coefficients in Riemannian calculus do \cite[Eq. (10.26)]{MTW};%
\begin{equation*}
\left( \Gamma {}^{\mu }{}_{\nu \rho }\right) ^{\prime }=\frac{\partial
x^{\prime \mu }}{\partial x^{\alpha }}\frac{\partial x^{\beta }}{\partial
x^{\prime \nu }}\frac{\partial x^{\gamma }}{\partial x^{\prime \rho }}\Gamma
^{\alpha }{}_{\beta \gamma }+\frac{\partial x^{\prime \mu }}{\partial
x^{\alpha }}\frac{\partial ^{2}x^{\alpha }}{\partial x^{\prime \nu }\partial
x^{\prime \rho }}.
\end{equation*}

\item Its invariance under local Lorentz transformations: The basis $\mathrm{%
s}_{\mu }$ and the gauge connection $\omega _{\mu }$ transform as $\mathrm{s}%
_{\mu }^{\prime }=\Lambda \mathrm{s}_{\mu }\overline{\Lambda }^{\ast }$ and $%
\omega _{\mu }^{\prime }=\Lambda \omega _{\mu }\overline{\Lambda }-\left(
\partial _{\mu }\Lambda \right) \overline{\Lambda }$, respectively, where $%
\overline{\Lambda }=\Lambda ^{-1}$. Using these relations and some of the
identities of Proposition \ref{Prop:CompAlgIds}, the claim follows by direct
calculation.
\end{itemize}

Note that $\Gamma {}^{\mu }{}_{\nu \rho }$, as defined by Eq. (\ref%
{Eq:MinimalConn}), need \textit{not} be symmetric in its lower two indices,
as the Christoffel coefficients are.

Because the metric transforms invariantly, compare Eq. (\ref{Eq:TransfMetric}%
), according to Eq. (\ref{Eq:CovDerScalarTensor}), the covariant derivative $%
D_{\rho }g_{\mu \nu }$ should equal $\nabla _{\rho }g_{\mu \nu }$. This is,
indeed, verified by the following explicit calculation, analogous to the
above proof of $0\equiv \nabla _{\rho }g_{\mu \nu }$, using Eqs. (\ref%
{Eq:CovDerQuatConj}), (\ref{Eq:CovDerBasis}), and (\ref{Eq:OmegaZeroScal}),
as well as some of the identities of Proposition \ref{Prop:CompAlgIds}:%
\begin{eqnarray*}
D_{\rho }g_{\mu \nu } &=&\left\langle D_{\rho }\mathrm{s}_{\mu },\mathrm{s}%
_{\nu }\right\rangle +\left\langle \mathrm{s}_{\mu },D_{\rho }\mathrm{s}%
_{\nu }\right\rangle \\
&=&\left\langle \nabla _{\rho }\mathrm{s}_{\mu }+\omega _{\rho }\mathrm{s}%
_{\mu }+\mathrm{s}_{\mu }\overline{\omega }_{\rho }^{\ast },\mathrm{s}_{\nu
}\right\rangle +\left\langle \mathrm{s}_{\mu },\nabla _{\rho }\mathrm{s}%
_{\nu }+\omega _{\rho }\mathrm{s}_{\nu }+\mathrm{s}_{\nu }\overline{\omega }%
_{\rho }^{\ast }\right\rangle \\
&=&\nabla _{\rho }\left\langle \mathrm{s}_{\mu },\mathrm{s}_{\nu
}\right\rangle +\left\langle \omega _{\rho }\mathrm{s}_{\mu },\mathrm{s}%
_{\nu }\right\rangle +\left\langle \mathrm{s}_{\mu }\overline{\omega }_{\rho
}^{\ast },\mathrm{s}_{\nu }\right\rangle +\left\langle \mathrm{s}_{\mu
},\omega _{\rho }\mathrm{s}_{\nu }\right\rangle +\left\langle \mathrm{s}%
_{\mu },\mathrm{s}_{\nu }\overline{\omega }_{\rho }^{\ast }\right\rangle \\
&=&\nabla _{\rho }\left\langle \mathrm{s}_{\mu },\mathrm{s}_{\nu
}\right\rangle +\left\langle \omega _{\rho },\mathrm{s}_{\nu }\overline{%
\mathrm{s}}_{\mu }\right\rangle +\left\langle \overline{\omega }_{\rho
}^{\ast },\overline{\mathrm{s}}_{\mu }\mathrm{s}_{\nu }\right\rangle
+\left\langle \mathrm{s}_{\mu }\overline{\mathrm{s}}_{\nu },\omega _{\rho
}\right\rangle +\left\langle \overline{\mathrm{s}}_{\nu }\mathrm{s}_{\mu },%
\overline{\omega }_{\rho }^{\ast }\right\rangle \\
&=&\nabla _{\rho }\left\langle \mathrm{s}_{\mu },\mathrm{s}_{\nu
}\right\rangle +\left\langle \omega _{\rho },\mathrm{s}_{\mu }\overline{%
\mathrm{s}}_{\nu }+\mathrm{s}_{\nu }\overline{\mathrm{s}}_{\mu
}\right\rangle +\left\langle \overline{\omega }_{\rho }^{\ast },\overline{%
\mathrm{s}}_{\mu }\mathrm{s}_{\nu }+\overline{\mathrm{s}}_{\nu }\mathrm{s}%
_{\mu }\right\rangle \\
&=&\nabla _{\rho }g_{\mu \nu }+2g_{\mu \nu }\left\langle 1,\omega _{\rho }+%
\overline{\omega }_{\rho }^{\ast }\right\rangle \\
&=&\nabla _{\rho }g_{\mu \nu }.
\end{eqnarray*}

\subsection{Various}

A final comment, before ending this section:\ Consider a quaternionic vector
field $V=V^{\mu }\mathrm{s}_{\mu }\in \mathbb{C}\otimes \mathbb{H}$, where $%
V^{\mu }\in \mathbb{C}$. The covariant derivative $D_{\mu }V$ is given by%
\begin{eqnarray*}
D_{\rho }V &=&\partial _{\rho }V+\omega _{\rho }V+V\overline{\omega }_{\rho
}^{\ast } \\
&=&\partial _{\rho }\left( V^{\mu }\mathrm{s}_{\mu }\right) +\omega _{\rho
}\left( V^{\mu }\mathrm{s}_{\mu }\right) +\left( V^{\mu }\mathrm{s}_{\mu
}\right) \overline{\omega }_{\rho }^{\ast } \\
&=&\left( \partial _{\rho }V^{\mu }\right) \mathrm{s}_{\mu }+V^{\mu }\left(
\partial _{\rho }\mathrm{s}_{\mu }+\omega _{\rho }\mathrm{s}_{\mu }+\mathrm{s%
}_{\mu }\overline{\omega }_{\rho }^{\ast }\right)  \\
&=&\left( \partial _{\rho }V^{\mu }\right) \mathrm{s}_{\mu }+\Gamma {}^{\mu
}{}_{\nu \rho }V^{\nu }\mathrm{s}_{\mu } \\
&=&\left( \partial _{\rho }V^{\mu }+\Gamma {}^{\mu }{}_{\nu \rho }V^{\nu
}\right) \mathrm{s}_{\mu },
\end{eqnarray*}%
using Eq. (\ref{Eq:CovDerVector}), and Eq. (\ref{Eq:MinimalConn}) in the
form $\Gamma {}^{\mu }{}_{\nu \rho }\mathrm{s}_{\mu }\equiv \partial _{\rho }%
\mathrm{s}_{\nu }+\omega _{\rho }\mathrm{s}_{\nu }+\mathrm{s}_{\nu }%
\overline{\omega }_{\rho }^{\ast }$. This may be written as%
\begin{eqnarray*}
\left\langle \mathrm{s}^{\mu },D_{\rho }V\right\rangle  &=&\partial _{\rho
}V^{\mu }+\Gamma ^{\mu }{}_{\nu \rho }V^{\nu }, \\
V^{\mu } &=&\left\langle \mathrm{s}^{\mu },V\right\rangle ,
\end{eqnarray*}%
which is completely analogous to the usual expression in Riemannian calculus
for a vector field $\mathbf{V}=V^{\mu }\mathbf{e}_{\mu }$;%
\begin{eqnarray*}
\left\langle \mathbf{\omega }^{\mu },\nabla _{\rho }\mathbf{V}\right\rangle 
&=&\partial _{\rho }V^{\mu }+\Gamma ^{\mu }{}_{\nu \rho }V^{\nu }, \\
V^{\mu } &=&\left\langle \mathbf{\omega }^{\mu },\mathbf{V}\right\rangle .
\end{eqnarray*}

\section{Field strength tensors}

In analogy with the expression $R^{\mu }{}_{\nu \rho \sigma }\equiv
\left\langle \mathbf{\omega }^{\mu },\left[ \nabla _{\rho },\nabla _{\sigma }%
\right] \mathbf{e}_{\nu }\right\rangle $ for the Riemannian curvature tensor
(in a coordinate basis) \cite[Sec. 11.3]{MTW}, the expression $\left\langle 
\mathrm{s}^{\mu },\left[ D_{\rho },D_{\sigma }\right] \mathrm{s}_{\nu
}\right\rangle $ is calculated, suspending for the moment the minimality
condition $0\equiv D_{\mu }\mathrm{s}_{\nu }$: Because $D_{\mu }\mathrm{s}%
_{\nu }$ is a quaternionic vector field, $D_{\rho }D_{\sigma }\mathrm{s}%
_{\mu }\equiv D_{\rho }\left( D_{\sigma }\mathrm{s}_{\mu }\right) $ is given
by%
\begin{eqnarray*}
D_{\rho }D_{\sigma }\mathrm{s}_{\mu } &=&\nabla _{\rho }\left( D_{\sigma }%
\mathrm{s}_{\mu }\right) +\omega _{\rho }\left( D_{\sigma }\mathrm{s}_{\mu
}\right) +\left( D_{\sigma }\mathrm{s}_{\mu }\right) \overline{\omega }%
_{\rho }^{\ast } \\
&=&\nabla _{\rho }\left( \nabla _{\sigma }\mathrm{s}_{\mu }+\omega _{\sigma }%
\mathrm{s}_{\mu }+\mathrm{s}_{\mu }\overline{\omega }_{\sigma }^{\ast
}\right) \\
&&+\omega _{\rho }\left( \nabla _{\sigma }\mathrm{s}_{\mu }+\omega _{\sigma }%
\mathrm{s}_{\mu }+\mathrm{s}_{\mu }\overline{\omega }_{\sigma }^{\ast
}\right) \\
&&+\left( \nabla _{\sigma }\mathrm{s}_{\mu }+\omega _{\sigma }\mathrm{s}%
_{\mu }+\mathrm{s}_{\mu }\overline{\omega }_{\sigma }^{\ast }\right) 
\overline{\omega }_{\rho }^{\ast },
\end{eqnarray*}%
which implies, various terms cancelling due to antisymmetry (and
associativity of the complexified quaternions),%
\begin{eqnarray*}
\left[ D_{\rho },D_{\sigma }\right] \mathrm{s}_{\mu } &=&\left[ \nabla
_{\rho },\nabla _{\sigma }\right] \mathrm{s}_{\mu } \\
&&+\left( \nabla _{\rho }\omega _{\sigma }-\nabla _{\sigma }\omega _{\rho }+ 
\left[ \omega _{\rho },\omega _{\sigma }\right] \right) \mathrm{s}_{\mu } \\
&&+\mathrm{s}_{\mu }\left( \nabla _{\rho }\overline{\omega }_{\sigma }^{\ast
}-\nabla _{\sigma }\overline{\omega }_{\rho }^{\ast }-\left[ \overline{%
\omega }_{\rho }^{\ast },\overline{\omega }_{\sigma }^{\ast }\right] \right)
,
\end{eqnarray*}%
which, using the identity $\left[ \overline{x}^{\ast },\overline{y}^{\ast }%
\right] \equiv -\overline{\left[ x,y\right] }^{\ast }$, may be rewritten as%
\begin{eqnarray*}
\left[ D_{\rho },D_{\sigma }\right] \mathrm{s}_{\mu } &=&\left[ \nabla
_{\rho },\nabla _{\sigma }\right] \mathrm{s}_{\mu }+\Omega _{\rho \sigma }%
\mathrm{s}_{\mu }+\mathrm{s}_{\mu }\overline{\Omega }_{\rho \sigma }^{\ast },
\\
\Omega _{\rho \sigma } &\equiv &\nabla _{\rho }\omega _{\sigma }-\nabla
_{\sigma }\omega _{\rho }+\left[ \omega _{\rho },\omega _{\sigma }\right] ,
\end{eqnarray*}%
with $\Omega _{\rho \sigma }$, seemingly, the field strength tensor
corresponding to the gauge connection $\omega _{\mu }$. This result is
analogous to the following result in General Relativity (augmented with a
vierbein $e^{a}{}_{\mu }$ and a spin connection $\omega _{\mu }{}^{a}{}_{b}$%
, compare \cite{Hooft}):%
\begin{eqnarray*}
\left[ D_{\rho },D_{\sigma }\right] e^{a}{}_{\nu } &=&-R^{\mu }{}_{\nu \rho
\sigma }e^{a}{}_{\mu }+F^{a}{}_{b\rho \sigma }e^{b}{}_{\nu }, \\
R^{\mu }{}_{\nu \rho \sigma } &\equiv &\partial _{\rho }\Gamma ^{\mu
}{}_{\nu \sigma }-\partial _{\sigma }\Gamma ^{\mu }{}_{\nu \rho }+\Gamma
^{\mu }{}_{\tau \rho }\Gamma ^{\tau }{}_{\nu \sigma }-\Gamma ^{\mu }{}_{\tau
\sigma }\Gamma ^{\tau }{}_{\nu \rho }, \\
F^{a}{}_{b\rho \sigma } &\equiv &\partial _{\rho }\omega _{\sigma
}{}^{a}{}_{b}-\partial _{\sigma }\omega _{\rho }{}^{a}{}_{b}+\omega _{\rho
}{}^{a}{}_{c}\omega _{\sigma }{}^{c}{}_{b}-\omega _{\sigma
}{}^{a}{}_{c}\omega _{\rho }{}^{c}{}_{b},
\end{eqnarray*}%
where $D_{\rho }e^{a}{}_{\nu }\equiv \partial _{\rho }e^{a}{}_{\nu }-\Gamma
^{\mu }{}_{\nu \rho }e^{a}{}_{\mu }+\omega _{\rho }{}^{a}{}_{b}e^{b}{}_{\nu
} $. Note that $\left\langle \mathrm{s}^{\mu },\left[ \nabla _{\rho },\nabla
_{\sigma }\right] \mathrm{s}_{\nu }\right\rangle \equiv R^{\mu }{}_{\nu \rho
\sigma }\left( \Gamma \right) $, by way of Eq. (\ref{Eq:NablaCovDerBasis}),
where, of course, the connection coefficients $\Gamma {}^{\mu }{}_{\nu \rho
} $ entering into $R^{\mu }{}_{\nu \rho \sigma }\left( \Gamma \right) $ are
those defined in Eq. (\ref{Eq:MinimalConn}). If $0\equiv D_{\rho }\mathrm{s}%
_{\nu }$, then%
\begin{equation}
0=\left[ \nabla _{\rho },\nabla _{\sigma }\right] \mathrm{s}_{\mu }+\Omega
_{\rho \sigma }\mathrm{s}_{\mu }+\mathrm{s}_{\mu }\overline{\Omega }_{\rho
\sigma }^{\ast }.  \label{Eq:FieldStrengthsRel}
\end{equation}

Decomposing the gauge connection as $\omega _{\mu }=\chi _{\mu }+\mathrm{i}%
gA_{\mu }$, where $\chi _{\mu }\in \mathbb{C}\otimes \mathrm{Vec}\left( 
\mathbb{H}\right) $ and $A_{\mu }\in \mathbb{R}$, in accordance with Eq. (%
\ref{Eq:OmegaZeroScal}), and $g\in \mathbb{R}$ is some coupling constant,
the field strength tensor decomposes as%
\begin{eqnarray*}
\Omega _{\mu \nu } &=&K_{\mu \nu }+\mathrm{i}gF_{\mu \nu }, \\
K_{\mu \nu } &\equiv &\nabla _{\mu }\chi _{\nu }-\nabla _{\nu }\chi _{\mu }+ 
\left[ \chi _{\mu },\chi _{\nu }\right] \in \mathbb{C}\otimes \mathrm{Vec}%
\left( \mathbb{H}\right) , \\
F_{\mu \nu } &\equiv &\nabla _{\mu }A_{\nu }-\nabla _{\nu }A_{\mu }\in 
\mathbb{R},
\end{eqnarray*}%
because $A_{\mu }\in \mathbb{R}$ commutes with any element of $\mathbb{C}%
\otimes \mathbb{H}$, and in particular with $\chi _{\mu }$ (and itself).
Note that $F_{\mu \nu }$ does \textit{not}, as usual in a coordinate basis
in Riemannian calculus, reduce to $\partial _{\mu }A_{\nu }-\partial _{\nu
}A_{\mu }$, because the connection coefficients $\Gamma {}^{\mu }{}_{\nu
\rho }$, as given by Eq. (\ref{Eq:MinimalConn}), are in general not
symmetric in the lower two indices.

\section{Unsettled issues}

Below are listed unsettled issues which, it is hoped, will be settled in the
near future, either by the author himself, or by some reader of this paper:

\begin{itemize}
\item Local $\mathrm{U}\left( 1\right) $ gauge transformation: It is quite
tempting to interprete $A_{\mu }$ and $F_{\mu \nu }$ as the electromagnetic
gauge connection and field strength tensor, respectively. This hypothesis is
strengthened by the fact that under the following local $\mathrm{U}\left(
1\right) $ gauge transformation,%
\begin{eqnarray}
\omega _{\mu }^{\prime } &=&\omega _{\mu }-\mathrm{i}\partial _{\mu }\phi ,
\label{Eq:OmegaU1} \\
\psi _{L}^{\prime } &=&\exp \left( +\mathrm{i}\phi \right) \psi _{L},
\label{Eq:PsiLU1} \\
\psi _{R}^{\prime } &=&\exp \left( -\mathrm{i}\phi \right) \psi _{R},
\label{Eq:PsiRU1}
\end{eqnarray}%
where $\phi \equiv \phi \left( x^{\mu }\right) \in \mathbb{R}$, consistent
with Eq. (\ref{Eq:OmegaZeroScal}), the covariant derivatives $D_{\mu }\psi
_{L}$ and $D_{\mu }\psi _{R}$ transform covariantly;%
\begin{eqnarray*}
\left( D_{\mu }\psi _{L}\right) ^{\prime } &=&\exp \left( +\mathrm{i}\phi
\right) \left( D_{\mu }\psi _{L}\right) , \\
\left( D_{\mu }\psi _{R}\right) ^{\prime } &=&\exp \left( -\mathrm{i}\phi
\right) \left( D_{\mu }\psi _{R}\right) .
\end{eqnarray*}%
Interestingly, the local gauge transformation given in Eq. (\ref{Eq:OmegaU1}%
) does not affect the covariant derivatives of any (tensor-valued)
quaternionic vector fields, because $\omega _{\mu }V_{\sigma _{1}\cdots
\sigma _{n}}^{\rho _{1}\cdots \rho _{m}}+V_{\sigma _{1}\cdots \sigma
_{n}}^{\rho _{1}\cdots \rho _{m}}\overline{\omega }_{\mu }^{\ast }$ in Eq. (%
\ref{Eq:CovDerVectorTensor}) is left unchanged due to the complex
conjugation in $\overline{\omega }_{\mu }^{\ast }$ [see also added note
below]. So, if the hypothesis of interpreting $A_{\mu }$ and $F_{\mu \nu }$
as the electromagnetic gauge connection and field strength tensor,
respectively, is correct, then at the level of (tensor-valued) complex
quaternionic fields it seems that only spinor fields can be electrically
charged. That, of course, flat out contradicts Nature, which does contain
charged fundamental vector particles, the weak bosons $W^{\pm }$. However,
it may be speculated that generalizing the formalism here presented from $%
\mathbb{C}\otimes \mathbb{H}$-valued fields to $\mathbb{C}\otimes \mathbb{O}$%
-valued fields, where $\mathbb{O}$ is the set of octonions \cite%
{Okubo,Springer and Veldkamp,Dundarer and Gursey}, which is indeed a very
natural thing to contemplate because the quaternions can be embedded into
the octonions in numerous ways, would give room for the existence of $W^{\pm
}$: for instance, it might be speculated that these bosons are a consequence
of interactions between a $\left( \mathbb{C}\otimes \mathbb{H}\right) $-part
of $\mathbb{C}\otimes \mathbb{O}$ and its complement $\left( \mathbb{C}%
\otimes \mathbb{O}\right) \backslash \left( \mathbb{C}\otimes \mathbb{H}%
\right) $. Also in favor of generalizing from quaternions to octonions is
the 'threeness' of the Fano plane, compare \cite[Fig. 1.1]{Okubo} or \cite[%
Sec. 2.1]{Baez}, of the seven imaginary units of the octonions which, as
also speculated by others, for instance \cite{Gunaydin and Gursey}, is quite
suggestive with respect to the threeness of color, or the threeness of the
family structure.

\item Potential symmetry of $\Gamma {}^{\rho }{}_{\mu \nu }$: In analogy
with the Christoffel coefficients of usual Riemannian calculus, it would be
desirable to have $\Gamma {}^{\rho }{}_{\mu \nu }=\Gamma {}^{\rho }{}_{\nu
\mu }$, because, for one thing, it would imply $\nabla _{\mu }A_{\nu
}-\nabla _{\nu }A_{\mu }\equiv \partial _{\mu }A_{\nu }-\partial _{\nu
}A_{\mu }$. By way of Eqs. (\ref{Eq:Nondegeneracy}) and (\ref{Eq:MinimalConn}%
), it is equivalent to%
\begin{equation*}
\partial _{\mu }\mathrm{s}_{\nu }+\omega _{\mu }\mathrm{s}_{\nu }+\mathrm{s}%
_{\nu }\overline{\omega }_{\mu }^{\ast }=\partial _{\nu }\mathrm{s}_{\mu
}+\omega _{\nu }\mathrm{s}_{\mu }+\mathrm{s}_{\mu }\overline{\omega }_{\nu
}^{\ast },
\end{equation*}%
which relates the gauge connection $\omega _{\mu }$ to the basis $\mathrm{s}%
_{\mu }$. As of yet, it has not been figured out how to solve this equation
for $\omega _{\mu }=\omega _{\mu }\left( \mathrm{s}_{\nu }\right) $.

\item Lagrangian: From Eq. (\ref{Eq:FieldStrengthsRel}) follows that%
\begin{equation*}
0=R^{\mu }{}_{\nu \rho \sigma }\left( \Gamma \right) +\left\langle \mathrm{s}%
^{\mu },\Omega _{\rho \sigma }\mathrm{s}_{\nu }+\mathrm{s}_{\nu }\overline{%
\Omega }_{\rho \sigma }^{\ast }\right\rangle ,
\end{equation*}%
where the connection coefficients $\Gamma {}^{\mu }{}_{\nu \rho }$ entering
into $R^{\mu }{}_{\nu \rho \sigma }\left( \Gamma \right) \equiv \left\langle 
\mathrm{s}^{\mu },\left[ \nabla _{\rho },\nabla _{\sigma }\right] \mathrm{s}%
_{\nu }\right\rangle $ are those defined in Eq. (\ref{Eq:MinimalConn}).
Because $\Omega _{\rho \sigma }\mathrm{s}_{\nu }+\mathrm{s}_{\nu }\overline{%
\Omega }_{\rho \sigma }^{\ast }\in \left( \mathbb{C}\otimes \mathbb{H}%
\right) ^{-}$, these components $R^{\mu }{}_{\nu \rho \sigma }\left( \Gamma
\right) $ are real-valued, which, of course, also is evident from the
real-valuedness of the connection coefficients $\Gamma {}^{\mu }{}_{\nu \rho
}$. On grounds of this relation, it may be speculated that a Lagrangian,
analogous to the Einstein-Hilbert Lagrangian, could be defined as (with $%
\kappa \in \mathbb{R}$ some constant),%
\begin{eqnarray*}
\frac{1}{\kappa }L &=&R^{\mu \nu }{}_{\mu \nu }\left( \Gamma \right) \\
&=&-\left\langle \mathrm{s}^{\mu },\Omega _{\mu \nu }\mathrm{s}^{\nu }+%
\mathrm{s}^{\nu }\overline{\Omega }_{\mu \nu }^{\ast }\right\rangle \\
&=&-\left\langle \mathrm{s}^{\mu }\overline{\mathrm{s}}^{\nu },\Omega _{\mu
\nu }\right\rangle -\left\langle \overline{\mathrm{s}}^{\nu }\mathrm{s}^{\mu
},\overline{\Omega }_{\mu \nu }^{\ast }\right\rangle \\
&=&-\left\langle \mathrm{s}^{\mu }\overline{\mathrm{s}}^{\nu },\Omega _{\mu
\nu }\right\rangle -\left\langle \mathrm{s}^{\nu }\overline{\mathrm{s}}^{\mu
},\overline{\Omega }_{\mu \nu }\right\rangle ^{\ast } \\
&=&-\left\langle \mathrm{s}^{\mu }\overline{\mathrm{s}}^{\nu },\Omega _{\mu
\nu }\right\rangle -\left\langle \mathrm{s}^{\mu }\overline{\mathrm{s}}^{\nu
},\Omega _{\mu \nu }\right\rangle ^{\ast } \\
&=&-2\func{Re}\left( \left\langle \mathrm{s}^{\mu }\overline{\mathrm{s}}%
^{\nu },\Omega _{\mu \nu }\right\rangle \right) ,
\end{eqnarray*}%
using $\overline{\mathrm{s}}_{\mu }^{\ast }=-\mathrm{s}_{\mu }$, and some of
the identities of Proposition \ref{Prop:CompAlgIds}. However, in regards to
the former hypothesis of interpreting $A_{\mu }$ and $F_{\mu \nu }$ as the
electromagnetic gauge connection and field strength tensor, respectively,
this Lagrangian seems less than optimal, because it does not give a
quadratic curvature term in $F_{\mu \nu }$. To obtain such a quadratic
Lagrangian, the Lagrangian could instead be considered defined as%
\begin{eqnarray*}
\frac{1}{\kappa }L &=&\func{Re}\left( \left\langle \Omega ^{\mu \nu },\Omega
_{\mu \nu }\right\rangle \right) \\
&=&\func{Re}\left( \left\langle K^{\mu \nu }+\mathrm{i}gF^{\mu \nu },K_{\mu
\nu }+\mathrm{i}gF_{\mu \nu }\right\rangle \right) \\
&=&\func{Re}\left( \left\langle K^{\mu \nu },K_{\mu \nu }\right\rangle
\right) -g^{2}F^{\mu \nu }F_{\mu \nu },
\end{eqnarray*}%
where the last equality follows from the fact that $\left\langle
x,y\right\rangle =0$, for any $x\in \mathbb{C}\otimes \mathrm{Scal}\left( 
\mathbb{H}\right) $ and $y\in \mathbb{C}\otimes \mathrm{Vec}\left( \mathbb{H}%
\right) $. This Lagrangian, however, introduces the new entity $\func{Re}%
\left( \left\langle K^{\mu \nu },K_{\mu \nu }\right\rangle \right) $:
whether or not it describes the gravitational force is as of yet completely
unsettled.
\end{itemize}

\textbf{Note added:} As commented shortly after Eqs. (\ref{Eq:OmegaU1})-(\ref%
{Eq:PsiRU1}) the covariant derivative of any (tensor-valued) quaternionic
vector field $V$ is invariant under a local $\mathrm{U}\left( 1\right) $
gauge transformation. This is consistent with, and could have been
anticipated from, the fact that $V^{\prime }=\Lambda V\overline{\Lambda }%
^{\ast }=V$ for $\Lambda =\exp \left( \mathrm{i}\phi \right) $. In fact,
local Lorentz transformations which obey $\overline{\Lambda }=\Lambda ^{-1}$%
, a relation frequently used in this paper, and local $\mathrm{U}\left(
1\right) $ gauge transformations $\Lambda =\exp \left( \mathrm{i}\phi
\right) $ can be treated in the following unified manner:

\begin{itemize}
\item Local invariance of the metric, Eq. (\ref{Eq:TransfMetric}): The
derivation holds also for a local $\mathrm{U}\left( 1\right) $ gauge
transformation, because $\left( \overline{\Lambda }\Lambda \right) ^{\ast
}=\left( \overline{\Lambda }\Lambda \right) ^{-1}$ for $\Lambda =\exp \left( 
\mathrm{i}\phi \right) $.

\item Transformation of the gauge connection, Eq. (\ref{Eq:TransfLorentzConn}%
): If $\overline{\Lambda }$ is replaced by $\Lambda ^{-1}$, thereby yielding%
\begin{equation*}
\omega _{\mu }^{\prime }=\Lambda \omega _{\mu }\Lambda ^{-1}-\left( \partial
_{\mu }\Lambda \right) \Lambda ^{-1},
\end{equation*}%
then Propositions \ref{Prop:CovDerSpinors} and \ref{Prop:CovDerVector} apply
to both local Lorentz transformations and local $\mathrm{U}\left( 1\right) $
gauge transformations. Note that this modified expression for $\omega _{\mu
}^{\prime }$ correctly reduces to Eq. (\ref{Eq:OmegaU1}) for $\Lambda =\exp
\left( \mathrm{i}\phi \right) $.
\end{itemize}

\section{Auxiliary material}

\subsection{Identities}

The following Proposition lists some useful identities for composition
algebras, a class to which the complexified quaternions belong, see for
instance \cite{Okubo} or \cite{Springer and Veldkamp}. Note that the
normalization of the inner products in \cite{Okubo} and \cite{Springer and
Veldkamp}, respectively, differ by a factor of $2$. The normalization used
in Eq. (\ref{Eq:ipDef}) is the normalization used in \cite{Okubo}.

\begin{proposition}
\label{Prop:CompAlgIds}The following identities hold for any composition
algebra:%
\begin{align}
\left\langle x,y\right\rangle & \equiv \left\langle y,x\right\rangle ,
\label{Eq:ipSym} \\
\left\langle x,y\right\rangle & \equiv \left\langle \overline{x},\overline{y}%
\right\rangle ,  \label{Eq:ipConj}
\end{align}%
and%
\begin{align}
\left\langle x,yz\right\rangle & \equiv \left\langle \overline{y}%
x,z\right\rangle ,  \label{Eq:ipMoveL} \\
\left\langle xy,z\right\rangle & \equiv \left\langle x,z\overline{y}%
\right\rangle .  \label{Eq:ipMoveR}
\end{align}
\end{proposition}

\subsection{Various}

\begin{proposition}
\label{Prop:RealValuedIP}Let $x,y\in \left( \mathbb{C}\otimes \mathbb{H}%
\right) ^{-}$. Then, $\left\langle x,y\right\rangle \in \mathbb{R}$.
\end{proposition}

\begin{proof}
Using $\overline{x}^{\ast }=-x$ and $\overline{y}^{\ast }=-y$:%
\begin{eqnarray*}
2\left\langle x,y\right\rangle ^{\ast } &=&\left( x\overline{y}+y\overline{x}%
\right) ^{\ast } \\
&=&x^{\ast }\overline{y}^{\ast }+y^{\ast }\overline{x}^{\ast } \\
&=&\overline{x}y+\overline{y}x \\
&=&2\left\langle x,y\right\rangle .
\end{eqnarray*}
\end{proof}

\end{document}